\documentclass[12pt]{aastex}
\usepackage{graphicx,bm}
\usepackage{txfonts}
\begin{document}
\title{Visible spectroscopy of 2003 UB$_{313}$: Evidence for N$_2$ ice on the
   surface of the largest TNO?}
\author{J. Licandro\altaffilmark{1,2}, W.M. Grundy\altaffilmark{3},
          N. Pinilla-Alonso\altaffilmark{4}, and P. Leisy\altaffilmark{1}}
\altaffiltext{1}{Isaac Newton Group, P.O.Box 321, E-38700, Santa Cruz de La Palma, 
Tenerife, Spain.}
\altaffiltext{2}{Instituto de Astrof\'{\i}sica de Canarias, c/V\'{\i}a L\'actea s/n, 
             E38205, La Laguna, Tenerife, Spain.}
\altaffiltext{3}{Lowell Observatory, 1400 West Mars Hill Road, Flagstaff, AZ
	     86001-4470.} 
\altaffiltext{4}{Fundaci\'on Galileo Galilei \& Telescopio 
        Nazionale Galileo, P.O.Box 565, E-38700, S/C de La Palma, Tenerife,
	     Spain.}
\email{licandro@ing.iac.es}


 
  \begin{abstract}
The recent discovery of two large trans-Neptunian objects (TNOs)
   2003 UB$_{313}$ and 2005 FY$_9$, with
   surface properties similar to those of Pluto, provides an
   exciting new laboratory for the study of processes considered for Pluto
   and Triton: volatile mixing and transport; atmospheric freeze-out and
   escape, ice chemistry, and nitrogen phase transitions.\\
   We studied the surface composition of TNO 2003 UB$_{313}$, 
   the first known TNO larger than Pluto.\\
   We report a visible spectrum  
   covering the 0.35-0.95$\mu$m  spectral 
   range, obtained with the 4.2m William Herschel Telescope 
   at ``El Roque de los 
   Muchachos" Observatory (La Palma, Spain). \\
   The visible spectrum of this TNO presents very prominent 
   absorptions bands 
   formed in solid CH$_4$.  At wavelengths shorter than 0.6 $\mu$m
   the spectrum 
   is almost featureless and slightly red (S'=4\%). 
   The icy-CH$_4$ bands 
   are significantly stronger than those of Pluto and slightly
   weaker than those observed in the spectrum of another giant
   TNO, 2005 FY$_9$, implying that 
   methane is more abundant on its surface than in Pluto's and close
   to that of the surface of 2005 FY$_9$. A shift of 15 $\pm$3 $\AA$ 
   relative to the position of the bands of the spectrum of
   laboratory CH$_4$ ice is observed
   in the bands at larger wavelengths (e.g. around 0.89 $\mu$m), but not 
   at shorter wavelengths (the band around 0.73 $\mu$m is not shifted) 
   this may be evidence for a vertical compositional gradient.  Purer methane 
   could have condensed first while 2003 UB$_{313}$ moved towards aphelion during
   the last 200 years, and as the atmosphere gradually collapsed, 
   the composition became more nitrogen-rich as the
   last, most volatile components condensed, and CH$_4$ diluted in N$_2$
   is present in the  outer surface layers.  
   \end{abstract}
%


%

\section{Introduction}

The trans-Neptunian 
region is populated by icy bodies (TNOs), remnant planetesimals from the 
early solar system formation stages (Edgeworth \cite{Edgeworth49}; 
Kuiper \cite{Kuiper51}). TNOs are the source of the short period comets
(Fern\'andez \cite{Fernandez80}) and are probably the most pristine 
objects in the Solar System. The low temperatures in this region ($\sim$40K), 
implies that
ices trapped at formation should be preserved and can provide key information 
on the composition and early conditions of the pre-solar nebula. 

The recent discovery of three very bright TNOs, 2003 EL$_{61}$ (Ortiz et
al.\ \cite{Ortiz2005}), 2003 UB$_{313}$ and 2005 FY$_9$ (Brown et al.\
\cite{Brownetal2005}, \cite{Brownetal2005b}) provides an excellent opportunity to 
obtain spectra of TNOs with sufficient S/N to obtain reliable mineralogical
information of their surfaces.

The spectra of three of the four largest members of the trans-neptunian belt, 
2003 UB$_{313}$, Pluto, and  2005 FY$_9$  are dominated by
strong methane ice absorption bands (Cruikshank et al.\ \cite{Cruik1993},
Brown et al.\ \cite{Brownetal2005}, Licandro et al.\ \cite{Licandroetal06})
Pluto also presents weak but unambiguous signatures of CO and 
N$_2$-ice (e.g. Cruikshank\ \cite{Cruik1998}). 
These bands are also detected in the spectrum of Neptune's satellite
Triton (Cruikshank et al.\ \cite{Cruik1993}), a possibly captured ex-TNO. 
The presence of frozen methane
on the surfaces of Pluto, Triton, 2005 UB${313}$ and 2005 FY$_9$ 
argues that the process
suggested by Spencer et al.\ (\cite{Spencer1997}) in which surface
methane is replenished from the interior, may be ubiquitous in large
trans-neptunian objects.  2005 FY$_9$ and 2003 UB$_{313}$ provide an
exciting new laboratory for the study of processes considered for Pluto
and Triton: volatile mixing and transport; atmospheric freeze-out and
escape, ice chemistry, and nitrogen phase transitions. In particular
the abundance of volatiles like CO and N$_2$ is important to determine
the possible presence of a bound atmosphere and constrain the formation conditions.

TNO 2005 UB$_{313}$ is the largest known object
in the trans-neptunian belt, with a surface albedo higher than that of Pluto
(2400+/-100 km or a size $\sim$5\% larger than Pluto, and p$_V$=86 $\pm$7 \%,
Brown et al.\ \cite{Brownetal06}). Discovered near aphelion at 97.50 AU, it will
take some 2 centuries to reach its perihelion at 38.2 AU. This huge 
variation in heliocentric distance causes large seasonal temperature
variations that should affect the sublimation
and recondensation of its surface volatiles.


In this paper
we present visible spectroscopy of 2005 UB$_{313}$
and compare it with spectra of Pluto and 2005 FY$_9$
(Licandro et al.\ \cite{Licandroetal06}) in order to
derive mineralogical information from its surface.


\section{Observations}

We observed 2003 UB$_{313}$ on 2005 October 20.03 UT 
with the 4.2m William Herschel telescope (WHT)
at the ``Roque de los Muchachos Observatory" (ORM, Canary Islands, Spain), 
under photometric conditions. 
The TNO had heliocentric distance 95.94 AU, geocentric 
distance 96.90 AU and phase angle 0.2$^o$. 

The visible spectrum (0.35-0.95$\mu$m) was obtained using 
the low resolution gratings (R300B in the blue arm, with a 
dispersion of 0.86$\AA/$pixel, and the R158R with a dispersion of 1.63 
$\AA$/pixel) of the double-armed spectrograph ISIS at WHT, and a 2" wide slit 
oriented at
the parallactic angle to minimize the spectral
effects of atmospheric dispersion. The tracking 
was at the TNO proper motion.  
Six 600s spectra were obtained by 
shifting the object by 10" in the slit to better correct the fringing. 
Calibration and extraction of the spectra were done using IRAF and
following  standard procedures  
(Massey et al. \cite{Massey1992}).
The six spectra of the TNO were averaged. The 
reflectance spectrum was obtained by dividing the spectrum of the TNO by the 
spectrum of the G2 star Landolt (SA) 93-101 (Landolt \cite{Landolt})
obtained the same night
just before and after the observation of the TNO at a similar 
airmass.

The final  reflectance spectrum, normalized at 0.6 $\mu$m is plotted in
Fig. 1 together with spectra of TNOs Pluto and 2005 FY$_9$. 
The spectrum of 2003 UB$_{313}$ presents all the 
methane ice absorption bands in this wavelength
range reported by Grundy et al.\ (\cite{GrundyCH4}),
even the weaker ones, and a slightly red slope, and  
it is very similar to the spectrum of Pluto and 2005~FY$_9$.



\section{Discussion}

The depth of the CH$_4$ ice absorption bands depends on
its abundance, texture, and/or the thickness of the methane-rich surface layer.
Licandro et al.\ (\cite{Licandroetal06}) noted that the near-infrared
spectrum of TNO 2005~FY$_9$ is very similar to the near-infrared
spectrum of 2003 UB$_{313}$ reported by Brown et al.\ (\cite{Brownetal2005d}).
The infrared bands in the spectrum of
2005~FY$_9$ are deeper than the same bands in Pluto's spectrum
(Licandro et al.\ \cite{Licandroetal06}),
which suggest that either the abundance of methane ice on the surface of 
2005~FY$_9$ is larger than on Pluto's surface, and/or the 
size of methane ice grains (or the thickness of the methane-rich surface 
layer)  is larger than that in Pluto's surface.
Unfortunately the low spectral resolution of the near-infrared
spectrum of 2005~FY$_9$,
and the S/N of the near-infrared spectrum of 2003 UB$_{313}$, 
do not permit
accurate measurements of band depths and central wavelengths
of the CH$_4$ bands.
Licandro et al.\ (\cite{Licandroetal06}) also reported that 
the prominent bands at 0.73$\mu$m and 0.89$\mu$m are 
$\sim$6 and $\sim$3 times deeper respectively in the spectrum of
2005~FY$_9$ than in Pluto's spectrum, while
bands in the near infrared spectrum are only  $<$2
times deeper, concluding
that light reflected from 2005 FY$_9$ samples
larger mean optical path lengths in CH$_4$ ice than light from Pluto
does. 

The depths of the CH$_4$ bands at 0.73$\mu$m and 0.89$\mu$m in the
spectrum of 2003 UB$_{313}$ are also greater than the same bands in 
the spectrum of Pluto (see Fig. \ref{spe2}), but slightly weaker
than those in the spectrum of 2005~FY$_9$. The 0.73$\mu$m and 0.89$\mu$m
bands are 1.9 and 1.1 times deeper, respectively, in the spectrum
of 2005~FY$_9$ than in the spectrum of 2003 UB$_{313}$. We conclude that light reflected
from 2003 UB313 requires mean optical path lengths in CH4 ice somewhere
between the values for Pluto and for 2005 FY9.
Compared with Pluto, larger grain sizes on the surface of
2003 UB$_{313}$ and 2005~FY$_9$ would accomplish this, 
as would higher CH$_4$ concentrations dissolved in nitrogen ice.  
Broader geographic distribution
of CH$_4$ ice on 2003 UB$_{313}$ and 2005 FY$_9$ could contribute as well, 
since Pluto's CH$_4$
ice is inhomogeneously distributed (Grundy \& Buie \cite{GruBuie01}).
Also the grain size and concentration of CH$_4$ seems to be larger
in 2005 FY$_9$ than in 2003 UB$_{313}$. 

%
%
%


In order to illustrate and  give support to the previous discussion, 
spectra of methane ice grains of different size were 
produced using the one-dimensional 
geometrical-optics formulation by Shkuratov et al.\ (\cite{Shkuratov}) 
and the optical constants of CH$_4$ ice from Grundy et al.\ \cite{GrundyCH4}, 
and compared with  
the spectra of 2003 UB$_{313}$, 2005 FY$_9$, and Pluto around
the 0.73 and 0.89 $\mu$m spectral bands (see Fig. \ref{spe2}).
With this comparison we do not pretend to produce mineralogical models 
of the surface of these TNOs.
The 0.73 $\mu$m band of 2003 UB$_{313}$, 2005 FY$_9$ and Pluto are closely
reproduced using 1.5, 4.5, and 0.5cm grains, respectively. The 0.89$\mu$m
band of 2003 UB$_{313}$, 2005 FY$_9$ and Pluto is better fitted with
1.5, 2.5 and 0.5cm grains respectively. 
As expected, larger grains are needed to reproduce the bands observed 
in the spectrum of 2005 FY$_9$ than  in the spectrum of 2003 UB$_{313}$
and both are larger than the grains used to reproduce the spectrum
of Pluto.
In the case of 2005 FY$_9$, smaller grains are needed to
reproduce the 0.89$\mu$m band than the 0.73$\mu$m band.
Licandro et al.\ (\cite{Licandroetal06}) found
that the weaker CH$_4$ bands at shorter wavelengths require
very large path lengths in CH$_4$ ice, since absorption by those
bands is much weaker than the stronger, near-infrared bands, which
require relatively little CH$_4$ to produce deep absorption bands.
Consequently, the shorter wavelengths are particularly sensitive to
regions having the most abundant CH$_4$ ice.
Different grain sizes are not used to reproduce the 0.73$\mu$m 
and 0.89$\mu$m bands observed in the spectrum of 2003 UB$_{313}$,
but notice that the fit of the 0.89$\mu$m  band is not as good as
that of the 0.73$\mu$m one. In particular, the 
center of the 0.89$\mu$m band is clearly shifter to shorter wavelengths 
relative to the modeled pure CH$_4$ ice spectrum. 


The shift of the CH$_4$ ice absorption bands relative to the wavelengths of 
pure methane ice absorption bands is another very important property
as can be indicative of dilution of  CH$_4$ in N$_2$ ice.
Pluto's CH$_4$ bands are seen to be partially shifted to shorter
wavelengths relative to the wavelengths of pure methane ice absorption
bands, indicating that at least some of the methane ice on Pluto's
surface is diluted in N$_2$ (Quirico et al.\ \cite{Quirico97}, Schmitt
et al.\ \cite{Schmitt98}, Dout\'e et al.\ \cite{Doute99}).  

Central wavelengths of the two deeper methane ice bands
in the visible spectrum of 2003 UB$_{313}$ were obtained by fitting a gaussian
around the bands, and are presented in Table~1. 
While the $3\nu_1 + 4\nu_4$ band is centered at  7296 $\AA$, very close
to the laboratory data, the $2\nu_1 + \nu_3 + 2\nu_4$ is at 8881 $\AA$
shifted by 16 $\AA$ from the position of pure methane ice.
To verify the wavelength callibration
of the spectrum, we measured the position of the bright sky lines, and
the uncertainties are smaller than 1 $\AA$. 
As the method used to determine the central wavelengths by fitting gaussians
depends on the
spectral region considered around the minimum, we also obtained the 
shifts by an auto-correlation against the model spectrum of pure
CH$_4$  in the spectral regions shown in Fig. \ref{spe2}. 
Shifts of -1 and 15 $\pm$3 $\AA$ were
obtained in the case of 2003 UB$_{313}$ for the 0.73 and 0.89 $\mu$m
bands respectively, while shifts of 2 and 5 $\pm$ 5 were obtained for 2005
FY$_9$. 

The band at 0.89 $\mu$m presents another characteristic that supports the
detection of CH$_4$ diluted in N$_2$ ice. In Fig. \ref{spe3} we present
the spectrum around the band and the spectrum of 
pure CH$_4$ shifted by 15 pixels. Notice
that the width of the band in the spectrum of 2003 UB$_{313}$ is
smaller than the width of the band in the spectrum of pure CH$_4$ ice.
This is what happens if the absorption is due to the monomer of CH$_4$ (Quirico
\& Schmitt \cite{Quirico97}) as in dilutions of CH$_4$ on N$_2$ at low
concentrations.
Brown et al.\ (\cite{Brownetal2005d}) measured the central waveleghts of several bands
in their near-infrared spectrum of 2003 UB$_{313}$ and compared them
to the position of pure methane at 30K and methane diluted in N$_2$ ice
from Quirico \& Schmitt (\cite{Quirico97}) laboratory measurements. They obtained
a mean shift of the four better defined methane bands of 15 $\pm$ 5 $\AA$ and
concluded that while a small amount of dissolved methane may be present, the
band positions suggest that the majority of methane is in essentially 
pure form. In the case of Pluto, Rudy et al.\ (\cite{Rudyetal03}) reported shifts in 
the near-infrared that are very similar to that in the 0.89$\mu$m band.
Considering the uncertainties, 
the shifts reported by Brown et al.\ (\cite{Brownetal2005d}) are not necessarily
discrepant
with our measurements. An unshifted methane band can correspond
either to pure methane ice or CH$_4$ diluted in N$_2$ ice at a relatively
high concentration (Quirico \& Schmitt 1997).

The shift observed at larger wavelengths, but not 
at shorter wavelengths, observed in the spectrum of 2003 UB$_{313}$, 
could be evidence for a vertical compositional 
gradient.
The weaker bands are formed on average more
deeply within the surface than the cores of the stronger bands are.  If
the weak bands look un-shifted and the strong bands look shifted, that
could indicate that purer methane condensed first, and, as the atmosphere
gradually collapsed while  2003 UB$_{313}$ moved towards aphelion
during the last two centuries, the composition became more nitrogen-rich as 
the last, most volatile components condensed.  N$_2$ is much more volatile than
CH$_4$ and so should survive in gaseous state to lower temperatures than
CH$_4$ would as 2003 UB$_{313}$  moves away from perihelion and cools.


CO and N$_2$ ices have indisputably detected in Pluto's spectrum
(Owen et al.\ \cite{Owen1993}). The hexagonal $\beta$ phase of N$_2$
ice was detected by means of its 2.15 $\mu$m absorption band and CO ice
was detected by means of a pair of narrow bands at 2.35 and 1.58 $\mu$m.
The spectral S/N of the spectrum of 2003 UB$_{313}$
(Brown et al.\ \cite{Brownetal2005d}) is not sufficient to see the CO absorptions.
The N$_2$ band would also be difficult to detect, even if N$_2$ were a
major component of the surface of 2003 UB$_{313}$, because the nitrogen
absorption has about a factor of a thousand smaller peak absorption
coefficient than that of the nearby CH$_4$ band at 2.2 $\mu$m, which
dominates that spectral region.  It is also
possible that surface temperatures on 2003 UB$_{313}$ might be below the
35.6~K transition temperature between the warmer $\beta$ phase of N$_2$
ice, and the colder, cubic $\alpha$ phase of N$_2$ ice, which has an
extremely narrow 2.15~$\mu$m absorption, which would be unresolved in
Brown et al.\ (\cite{Brownetal2005d}) data (e.g., Grundy et al.\ \cite{GrundyN2}).  
Future, higher spectral
resolution observations will put more constraints on the presence of N$_2$
and CO ice on the surface of 2003 UB$_{313}$.

A final important characteristic of the spectrum is its colour.
The surface of 2003 UB$_{313}$ is slightly red.
To compare with Pluto we computed the ratio of
the reflectance spectrum at 0.825 and 0.590 $\mu$m as in Grundy \& Fink
(\cite{GruFink}). The value of this ratio is 1.10, and corresponds
to a spectral slope S'=4 \%/1000$\AA$. Pluto and 2005 FY$_9$ present a
slightly redder spectrum, with a ratio of 1.20 and 1.21 respectively
(S'=8.8 and 8.9 \%/1000$\AA$, Licandro et al.\ \cite{Licandroetal06}) .  
The most accepted hypothesis to explain the red colour of Pluto
is the existence of
complex organics molecules (tholins) formed from simple organics by
photolysis (e.g. Khare et al.\ \cite{Khare1984}). Thus tholins should be
less abundant in 2003 UB$_{313}$ than in Pluto and 2005 FY$_9$.


\section{Conclusions}

We present a new 0.35-9.4$\mu$m spectrum of the TNO 2003 UB$_{313}$. 
The spectrum is very similar to that of Pluto, with prominent CH$_4$ ice 
absorptions bands.  
At wavelengths $<0.6 \mu$m the spectrum is almost featureless and slightly 
red (S'=4$\pm$1\%/1000$\AA$) supporting the existence of
complex organics molecules (tholins) on its surface. 
The visible spectrum of 2003 UB$_{313}$
is not as red as spectra of Pluto and 2005 FY$_9$ (S'=8.8 and 
8.9 \%/1000$\AA$ respectively), 
thus complex organics should be less abundant on the surface of
2003 UB$_{313}$ than on the surfaces of Pluto and 2005 FY$_9$. 

The CH$_4$ ice bands in this new giant TNO are significantly stronger than 
those of Pluto, but weaker than those observed in the spectrum
of 2005 FY$_9$ (Licandro et al.\ \cite{Licandroetal06}).
Methane is more abundant and/or the methane ice grain particles
(or the thickness of the surface ice layer)
are larger on
its  surface than on the surface of Pluto, and less abundant or 
composed of smaller grains
than on the surface of 2005 FY$_9$. 

A 15 $\AA$ shift of the central wavelength of the 0.89$\mu$m band relative to 
the pure methane band observed in the laboratory is observed. This shift
is indicative of the presence of methane diluted in N$_2$. On the other hand, 
the 0.73 $\mu$m band is not significantly shifted.
This could be evidence for a vertical
compositional gradient consistent with purer methane condensing first,
with the composition becoming more nitrogen-rich as the last, most
volatile components of the atmosphere condensed.  Such a compositional
gradient could also arise via the solar gardening mechanism discussed
by Grundy \& Stansberry (\cite{GruStan00}).

{\bf Acknowledgements}
W.M. Grundy gratefully acknowledges support from NASA Planetary Geology
\&\ Geophysics grant NNG04G172G.

\begin{table}
\centering
\begin{tabular}{c c c c c }
\hline
\hline
band & methane & 2003 UB$_{313}$  & 2005 FY$_9$  & Pluto  \\ 
  & ($\AA$) &  ($\AA$) &  ($\AA$) & ($\AA$) \\ \hline 
$3\nu_1+4\nu_4$	  & 7299 & 7296 & 7296 & 7290 \\ \hline 
$2\nu_1+\nu_3+2\nu_4$ & 8897 & 8881	& 8891 & 8885 \\ \hline 
\end{tabular} 
\caption{
Position of the prominent methane lines in the spectra of 
2005 UB$_{313}$, 2005 FY$_9$, and
Pluto. Laboratory data from Grundy et al.\
\cite{GrundyCH4}, Pluto data (with uncertainties $\sim$10\AA) from Grundy
\&\ Fink \cite{GruFink}, 2005 FY$_9$ data (with uncertainties $\sim$4\AA) from
Licandro et al. \cite{Licandroetal06}. }
\end{table}

   \begin{figure}
   \centering
   \includegraphics[angle=270,width=\columnwidth]{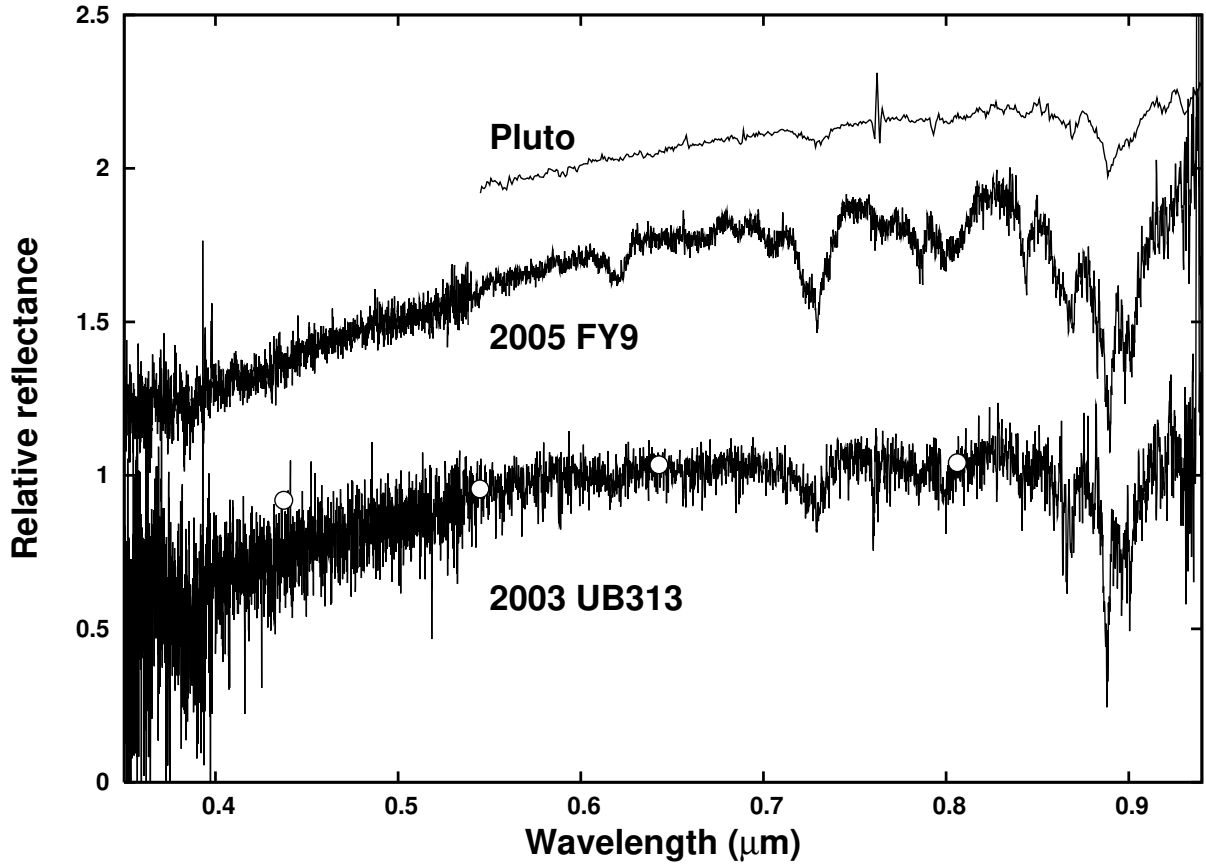}
   \caption{Reflectance spectra of 2003 UB$_{313}$ obtained on 
   2005 October 20.03 UT ,
   normalized at 0.6$\mu$m.
   The spectrum
   of Pluto (Grundy \&\ Fink \cite{GruFink})
   and the spectrum 2005 FY$_9$ (Licandro et al.\ \cite{Licandroetal06}), 
   both shifted vertically, are
   plotted for comparison.  Open circles are the relative 
   reflectance derived from $BVRI$ photometry reported by 
   Brown et al.\ (\cite{Brownetal2005d}).
   The spectrum of 2003 UB$_{313}$ is very similar
   to that of Pluto and 2005 FY$_9$, and reveals important features: 
   (1) the slope of the
   continuum in the visible range is slightly red, but less so than 
   those of Pluto and 2005 FY$_9$, indicative of the presence of 
   complex organics; (2) there are several CH$_4$ ice absorption bands;
   (3) The methane ice absorption bands in the spectrum of 2003 UB$_{313}$
   are deeper than the same bands in Pluto's spectrum but slightly weaker
   than the same bands in the spectrum of 2005 FY$_9$.}
              \label{spe1}
    \end{figure}

   \begin{figure}
   \centering
   \includegraphics[angle=270,width=110mm]{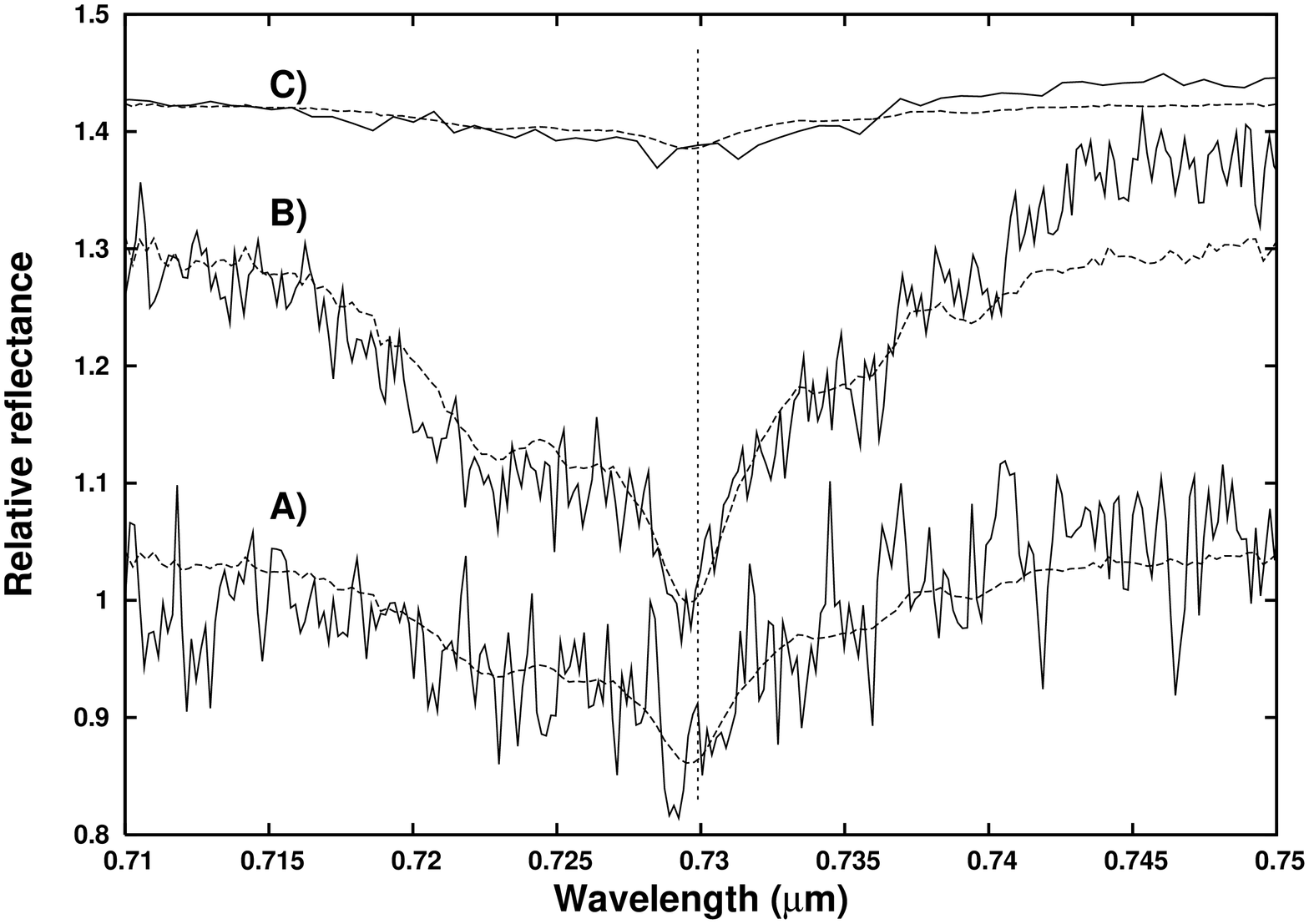}
   \includegraphics[angle=270,width=110mm]{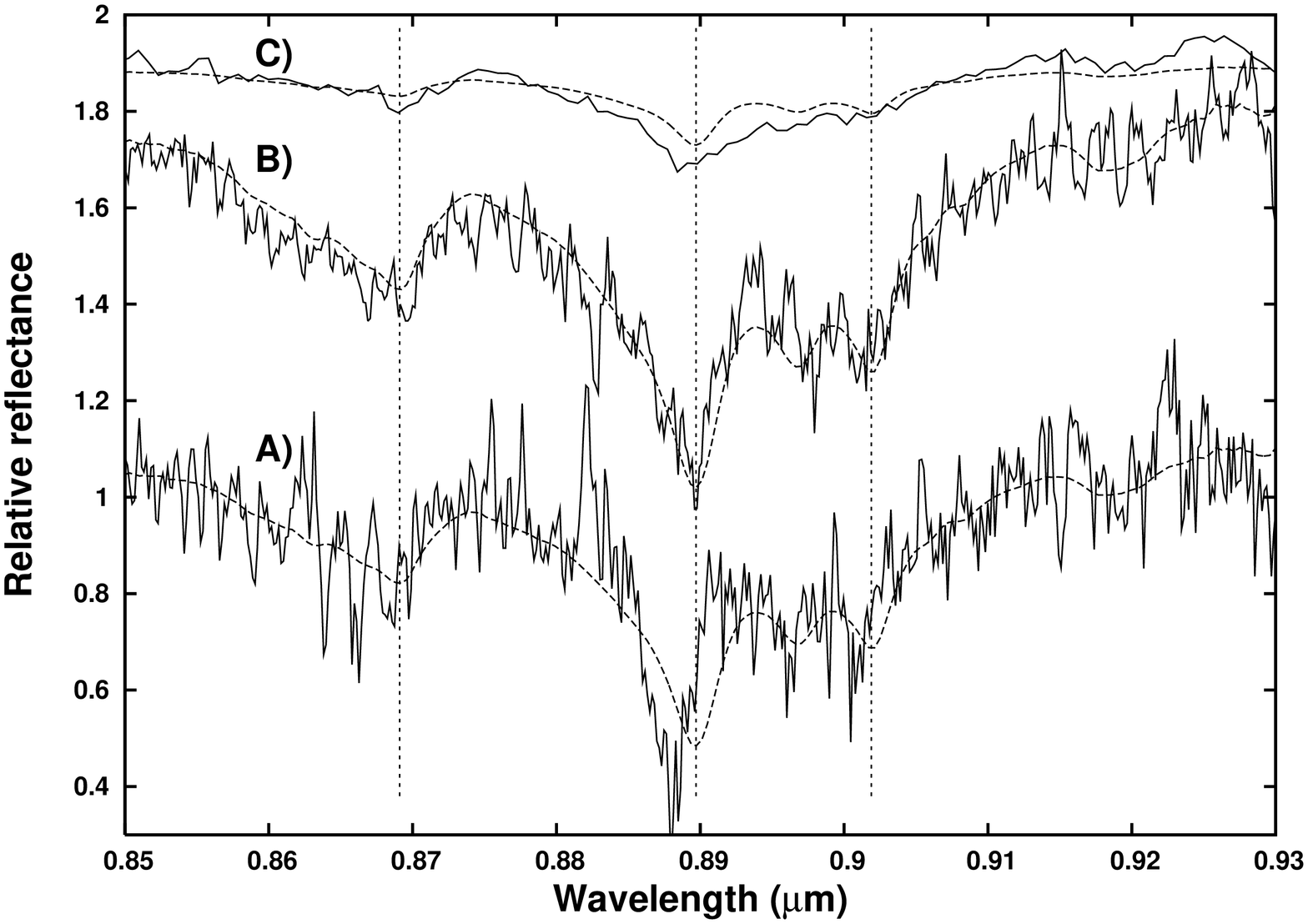}
   \caption{Reflectance spectra of TNOs 2003 UB$_{313}$, 2005 FY$_9$,
   and Pluto shifted vertically, in the two wavelength regions 
   of the most prominent
   $CH_4$ ice absorption bands.
   $Upper Figure$: (A) is the spectrum
   of 2003 UB$_{313}$ and overplotted (dashed lines) is the spectrum of pure
   methane ice grains of 1.5 cm diameter; (B) is the spectrum of 2005 FY$_9$
   and overplotted (dashed lines) the spectrum of pure
   methane ice grains of 4.5 cm diameter; (C) is the spectrum of Pluto and
   overplotted (dashed lines) the spectrum of pure
   methane ice grains of 500 $\mu$m diameter; vertical dashed lines 
   indicate the central position of pure
   methane ice bands (Grundy et al.\ \cite{GrundyCH4}).
   $Lower Figure$: (A) same as in upper figure; (B) is the spectrum of 
   2005 FY$_9$
   and overplotted (dashed lines) the spectrum of pure
   methane ice grains of 2.5 cm diameter; (C) as in upper figure; vertical
   dashed lines as in upper figure.}
              \label{spe2}
    \end{figure}

   \begin{figure}
   \centering
   \includegraphics[angle=270,width=\columnwidth]{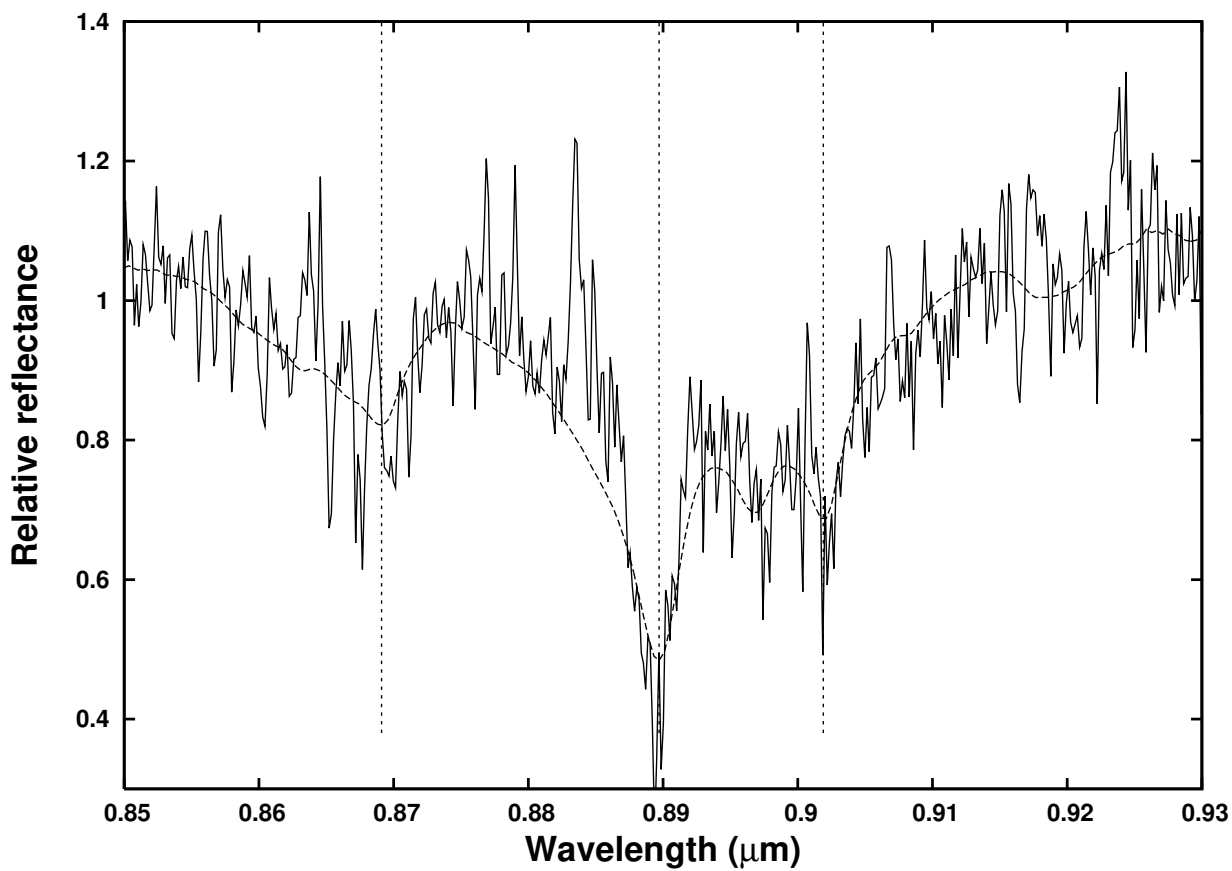}
   \caption{Reflectance spectrum of TNOs 2003 UB$_{313}$ (solid line) 
   and the spectrum of pure
   methane ice grains of 1.5 cm diameter shifted by 15 $\AA$ (dashed line).
   }
              \label{spe3}
    \end{figure}

%


\end{document}